\documentclass[12pt,fleqn]{article}

\usepackage{latexsym}
\usepackage{amsmath}
\usepackage{amssymb}

\renewcommand{\[}{$$}
\renewcommand{\]}{$$%
\par  \noindent  \hspace{-0.4em}}

\newcommand{\rf}[1]{(\ref{#1})}

\newcommand{\ba}{\begin{array}}
\newcommand{\ea}{\end{array}}

\newcommand{\be}{\begin{equation}}
\newcommand{\ee}{\end{equation}}

\newcommand{\ods}{\par \vspace{0.5cm} \par}

\newcommand{\const}{{\rm const}}

\newcommand{\dis}{\displaystyle }

\newcommand{\R}{{\mathbb R}}
\newcommand{\T}{{\mathbb T}}
\newcommand{\Z}{{\mathbb Z}}

\newcommand{\e}{{\bf e}}

\newcommand{\m}{\left( \ba{c}}
\newcommand{\ema}{\ea \right)}
\newcommand{\mm}{\left( \ba{cc}}

\newcommand{\no}{{\noindent}}

\newenvironment{Proof}{\noindent \small {\it Proof:\ }}%
{\normalsize \hfill {\mbox{$\Box$}} \par  \vspace{1.5ex}}

\newtheorem{prop}{Proposition}

\newtheorem{lem}[prop]{Lemma}

\newtheorem{cor}[prop]{Corollary}

\numberwithin{equation}{section}
\numberwithin{prop}{section}

\begin{document}

\title{\bf The sine-Gordon equation \\ on time scales }
\author{Jan L.\ Cie\'sli\'nski\thanks{e-mail:
\tt janek\,@\,alpha.uwb.edu.pl}, \ \ 
 Tomasz Nikiciuk\thanks{e-mail:
\tt niki\,@\,alpha.uwb.edu.pl} 
\\ {\footnotesize Uniwersytet w Bia\l ymstoku,
Wydzia\l \ Fizyki,}
\\ {\footnotesize ul.\ Lipowa 41, 15-424 Bia\l ystok, Poland} \\[2ex]
Kamil Wa\'skiewicz\thanks{e-mail:
\tt kamilwas@igf.edu.pl}, 
\\ {\footnotesize  Instytut Geofizyki PAN,} \footnotesize \\ { \footnotesize ul.\ Ks.\ Janusza 64, 01-452 Warszawa, Poland}
}

\date{}

\maketitle

\begin{abstract}
We formulate and discuss integrable  analogue of the sine-Gordon equation on arbitrary time scales. This unification contains the sine-Gordon equation, discrete sine-Gordon equation and the Hirota equation (doubly discrete sine-Gordon equation) as special cases. We present the Lax pair, check compatibility conditions and construct the Darboux-B\"acklund transformation. Finally, we obtain a soliton solution on arbitrary time scale. The solution is expressed by the so called Cayley exponential function. 
\end{abstract}


   \par


\ods

\section{Introduction}

The sine-Gordon equation \  $\phi,_{xy} = \sin\phi$  \  is one of the  classical soliton equations with numerous applications in many fields \cite{BEMS,Dav,Lam}, from differential geometry \cite{BP,Bour,MS,PM}  to applied physics, including  relativistic field theory \cite{Raj}, Josephson junctions \cite{Rem,Sco-junct}, propagation of deformations along DNA double helix \cite{Ya}, dispalcements in crystals \cite{FK}, domain walls in ferroelectric and ferromagnetic materials \cite{KIK}, 
mechanical transmission lines \cite{Rem,Sco}, and many others. 
In spite of the long history the sine-Gordon equation and its numerous extensions and generalizations still attracts attention of researchers \cite{DCW,FZ,LF,Sa,WRG}. In this paper we formulate and study  an integrable  extension of  the sine-Gordon equation on arbitrary time scales, which includes, as special cases, the discrete case  \cite{PL} and doubly discrete case (the Hirota equation)  \cite{Hir}. 

A time scale $\T$ is just any (non-empty) closed subset of $\R$  \cite{Hi}, including special cases like $\T = \R$, \ $\T= \Z$, \ $\T = h \Z$ and  $q$-calculus.   Time scales (or measure chains) were introduced in order to unify continuous and discrete calculus \cite{Hi2}.  We recall several notions which will be used throughout the paper (here we assume $t \in \T$, note that in the rest of the paper we have two time scales with elements denoted respectively by $x, y$).
The forward jump operator is defined as
\be
   t^\sigma := \inf \{ s \in \T : s > t \} .
\ee
we denote also 
\be
    \phi^\sigma (t) := \phi (t^\sigma) .
\ee
In the case $\T= h \Z$ the forward jump is a shift.  Graininess is defned by 
\be
\mu (t) := t^\sigma  - t  \ . 
\ee 
In the case $\T= h \Z$  we have $\mu = h$ (time step). In the case $\T=\R$, we have, obviously, $\mu=0$.  The delta derivative is defined by
\be
f^\Delta (t ) : = \lim_{ s \rightarrow t } \frac{ f ( t^\sigma  ) - f ( s )}{t^\sigma  - s} \ .
\ee
Note that 
\be \label{useful}
   f + \mu f^{\Delta} = f^{\sigma} .
\ee
Delta exponential function $e_a (t)$, see \cite{BP-exp,Hi},  satisfies initial value problem
\be   \label{deltaexp}
e_a^{\Delta} =  a e_a , \qquad  e_a (0) = 1 , 
\ee
where, in general, $a = a(t)$. We also have:
\be
     e_a^{\sigma} = (1 + a \mu) e_a 
\ee
In some applications (e.g., in trigonometry) another definition of the exponential function is much more convenient \cite{Ci-DCDS,Ci-JMAA}. This is the Cayley exponential function which satisfies  
\be  \label{Cayley}
    E_a^{\Delta} = \frac{1}{2} a (E_a + E_a^{\sigma}) , \qquad  E_a (0) = 1 ,   \qquad 
  E_a^{\sigma} = \frac{1 + \frac{a \mu}{2}}{1 - \frac{a \mu}{2}} E_a \ . 
\ee
In the continuous case ($\T=\R$)  exponential functions become identical and for $a=\const$ we have: $e_a (t) = E_a (t)$ $ = e^{a t}$. 

Dynamic equations on time scales are counterparts of differential equations. They unify continuous and discrete dynamical systems  \cite{AM,BP-I,Jac}. Some soliton equations on time scales were already studied within the Hamiltonian framework \cite{BSS,GGS,SBS}. This paper is a continuation of \cite{Ci-pseudo}, where the Darboux-B\"acklund transformation for a class of  $2\times 2$ linear problems was constructed. We will  construct this transformation for the integrable analogue of the sine-Gordon equation on time scales and compute a soliton solution expressed by the Cayley exponential function. 

\section{Lax pair}

A standard  Lax pair for the sine-Gordon equation $\phi,_{xy} = \sin\phi$: 
\be  \ba{l} \dis   \label{Lax-SG}
\Psi,_x = \mm i \zeta & - \frac{1}{2} \phi,_x  \\  \frac{1}{2} \phi,_x  & - i \zeta \ema \equiv 
i \zeta \sigma_3  - \frac{1}{2} i \phi,_x  \sigma_2 \ , \\[3ex]\dis 
\Psi,_y = \frac{1}{4 i \zeta} \mm \cos \phi  &  \sin \phi \\ \sin \phi & - \cos\phi \ema  \equiv 
 \frac{1}{4 i \zeta} \left( \cos\phi \sigma_3 + \sin\phi \sigma_1 \right) \ ,
\ea \ee 
where $\sigma_1, \sigma_2, \sigma_3$ are Pauli matrices. 
We postulate the following ime scale analogue for  this Lax pair:
\be   \ba{l} \dis   \label{Lax-time}
\Psi^{\Delta_x} = U \Psi \ ,   \quad U = i \zeta \sigma_3 - \frac{\sin \frac{\mu_x \phi^{\Delta_x}}{2}}{\mu_x} i \sigma_2  - \frac{1 - \cos \frac{\mu_x \phi^{\Delta_x}}{2} }{\mu_x} \\[3ex] \dis
\Psi^{\Delta_y} = V \Psi \ ,\quad V = \frac{1}{4 i\zeta} \left(   \cos \frac{\phi^{\sigma_y} + \phi}{2} \ \sigma_3 + \sin \frac{\phi^{\sigma_y} + \phi}{2} \ \sigma_1  
\right) 
\ea 
\ee
We make important assumption throughout this paper:
\be
   \mu_x = \mu_x (x) \ , \qquad  \mu_y = \mu_y (y) \ ,
\ee
i.e., we consider  time scales  \ ${\mathbb T}  = {\mathbb T}_1 \times {\mathbb T}_2$ with graininess $\mu = (\mu_x(x), \mu_y(y))$.   If $\mu  \rightarrow (0,0)$, then the Lax pair \rf{Lax-time} becomes \rf{Lax-SG}. 
It is convenient to denote (compare \cite{BP}): 
\be  \label{aldel}
  \delta := \frac{1}{2}  \phi^{\Delta_x}  \ , \qquad  \alpha := \frac{1}{2} \left(  \phi + \phi^{\sigma_y} \right) \ .
\ee
Then, the Lax pair \rf{Lax-time} takes the form
\be \ba{l} \dis  \label{Lax-aldel} 
\Psi^{\Delta_x } = \left( -  \zeta \e_3   +  \frac{\sin\mu_x\delta}{\mu_x}   \e_2 - \frac{1-\cos\mu_x\delta}{\mu_x}  \right)  \Psi \ ,  \\[3ex]   \dis
\Psi^{\Delta_y} =   \frac{1}{4 \zeta} \left( \e_3   \cos\alpha  + \e_1   \sin\alpha  \right)  \Psi ,
\ea \ee
where $\e_1 = - i \sigma_1, \ \e_2 = -  i \sigma_2, \  \e_3 = - i \sigma_3$. We have 
\be \ba{l}   \label{Cliff}
   \e_1 \e_2 = - \e_2 \e_1 = \e_3 , \qquad \e_2 \e_3 = - \e_3 \e_2 = \e_1 , \\[1ex] 
\e_3 \e_1 = - \e_1 \e_3 = \e_2 , \qquad    \e_1^2 = \e_2^2 = \e_3^2 = - 1 . 
\ea \ee
\ods

\begin{lem}  \label{lem-aldel}
  \be
  \alpha^{\Delta_x}  = \delta^{\sigma_y} + \delta \ , \qquad  \mu_x ( \delta + \delta^{\sigma_y} ) = \alpha^{\sigma_x} - \alpha .
\ee
\end{lem}

\begin{Proof} Using \rf{aldel} we get
\be
\alpha^{\Delta_x}  = \frac{1}{2} \left(   \phi^{\Delta_x} + \phi^{\sigma_2 \Delta_x}   \right) = \delta + \delta^{\sigma_y}  \ .
\ee
The second equality follows by \rf{useful}.   \end{Proof}

\begin{prop}
Compatibility conditions, $\dis U^{\Delta_y} - V^{\Delta_x} + U^{\sigma_y} V - V^{\sigma_x} U = 0 $,  for \rf{Lax-time} are given by 
\be  \label{dsG}  
 \frac{4}{\mu_x \mu_y} \sin \frac{\mu_x \mu_y \, \phi^{\Delta_x \Delta_y} }{4} = \sin \frac{\phi^{\sigma_x \sigma_y} + \phi^{\sigma_x} + \phi^{\sigma_y} + \phi}{4} \ , \\[1ex]
\ee
where  for $\mu_x =0$ or $\mu_y=0$  the left-hand side becomes $\phi^{\Delta_x \Delta_y}$  (as limits $\mu_x \rightarrow 0$ or $\mu_y \rightarrow 0$ suggest). 
\end{prop}

\begin{Proof}  Compatibility conditions yield a system of four scalar equations given by the decomposition in the basis $1, \e_1, \e_2, \e_3$. Coefficients by $\e_1, \e_3$ yield
 \be \ba{l}   \label{1-3}
(\sin\alpha)^{\Delta_x} = \frac{ \sin\mu_x \delta^{\sigma_y} \cos\alpha - (1-\cos\mu_x \delta^{\sigma_y}) \sin\alpha + \cos\alpha^{\sigma_x} \sin\mu_x \delta +  (1-\cos\mu_x \delta) \sin\alpha^{\sigma_x} }{\mu_x}   ,  \\[2ex]
(\cos\alpha)^{\Delta_x} = \frac{ - \sin\mu_x \delta^{\sigma_y} \sin \alpha - (1-\cos\mu_x \delta^{\sigma_y}) \cos\alpha -  \sin\alpha^{\sigma_x} \sin\mu_x \delta +  (1-\cos\mu_x \delta) \cos\alpha^{\sigma_x} }{\mu_x} .
\ea \ee
If $\mu_x \neq 0$, then we use  \rf{useful} to  transform these equations into
\be \ba{l}
\sin\alpha^{\sigma_x} \cos\mu_x\delta - \cos\alpha^{\sigma_x} \sin\mu_x\delta = \sin\alpha \cos\mu_x \delta^{\sigma_y}+ \cos\alpha \sin\mu_x \delta^{\sigma_y} , \\[1ex]
\cos\alpha^{\sigma_x} \cos\mu_x\delta + \sin\alpha^{\sigma_x} \sin\mu_x\delta = \cos\alpha \cos\mu_x \delta^{\sigma_y} - \sin\alpha \sin\mu_x \delta^{\sigma_y} .
\ea \ee 
Using well known trigonometric identities, we get 
\be \ba{l}  
\sin (\alpha^{\sigma_x} - \mu_x\delta)  = \sin (\mu_x \delta^{\sigma_y} + \alpha)   , \\[1ex]
\cos ( \alpha^{\sigma_x} - \mu_x \delta)  = \cos (\mu_x \delta^{\sigma_y}+\alpha)  ,
\ea \ee 
which is identically satisfied by \rf{useful}  and Lemma~\ref{lem-aldel}. If $\mu_x = 0$, then \rf{1-3} reduce to
\be \ba{l}
 \alpha^{\Delta_x} \cos\alpha =  (\delta^{\sigma_y} + \delta) \cos\alpha  , \qquad 
 \alpha^{\Delta_x} \sin\alpha =  (\delta^{\sigma_y} + \delta) \sin\alpha ,
\ea \ee
which is satisfied by  Lemma~\ref{lem-aldel}, as well.

Dealing with coefficients by $1$ and $\e_2$  we will  consider three cases separately. In the first case we assume  $\mu_x \neq 0$, $\mu_y \neq 0$.  Then  coefficients by $1$ and $\e_2$ yield  
\be \ba{l} 
\cos \mu_x \delta^{\sigma_y} = \cos \mu_x\delta + \frac{1}{4} \mu_x \mu_y   (\cos\alpha^{\sigma_x} - \cos\alpha)  , \\[2ex]
\sin  \mu_x \delta^{\sigma_y} = \sin \mu_x \delta + \frac{1}{4} \mu_x \mu_y   (\sin\alpha^{\sigma_x} + \sin\alpha) , 
\ea \ee 
Hence, using  trigonometric identities, we get 
\be \ba{l}   \label{sys01}
\sin \frac{ \mu_x (\delta^{\sigma_y} + \delta)}{2} \sin \frac{ \mu_x ( \delta^{\sigma_y} - \delta)}{2}  = \frac{1}{4} \mu_x \mu_y   \sin \frac{ \alpha^{\sigma_x} - \alpha}{2} \sin  \frac{ \alpha^{\sigma_x} + \alpha}{2}   , \\[2ex] 
\cos \frac{  \mu_x (\delta^{\sigma_y} + \delta)}{2} \sin \frac{ \mu_x (\delta^{\sigma_y} - \delta) }{2}  = \frac{1}{4} \mu_x \mu_y   \cos \frac{ \alpha^{\sigma_x} - \alpha}{2} \sin \frac{ \alpha^{\sigma_x} + \alpha}{2}   .
\ea \ee 
By virtue od Lemma~\ref{lem-aldel}  these equations reduce to the following single equation
\be
 \sin \frac{  \mu_x (\delta^{\sigma_y} - \delta) }{2} = \frac{1}{4} \mu_x \mu_y   \sin \frac{ \alpha^{\sigma_x} + \alpha}{2}  \ .
\ee
Using \rf{aldel}  and taking into account that $\delta^{\sigma_y} = \delta + \mu_y \delta^{\Delta_y}$ we get \rf{dsG}.

In the second case  we assume $\mu_x \neq 0$ and $\mu_y = 0$ (hence, in particular, $\delta^{\sigma_y} = \delta$ and $\alpha = \phi$). Then coefficients by $1$ and $\e_2$ yield  
\be \ba{l} 
- \delta^{\Delta_y}  \sin \mu_x \delta =   \frac{1}{4}    (\cos\alpha^{\sigma_x} - \cos\alpha)  , \\[2ex]
\delta^{\Delta_y}  \cos\mu_x \delta =   \frac{1}{4}     (\sin\alpha^{\sigma_x} + \sin\alpha) ,.
\ea \ee 
In this case Lemma~\ref{lem-aldel} implies $\mu_x \delta = \frac{1}{2} (\alpha^{\sigma_x} - \alpha)$.  Therefore
\be \ba{l} 
\sin \frac{ \alpha^{\sigma_x} - \alpha}{2} \left(  \delta^{\Delta_y}  -  \frac{1}{2}    \sin \frac{ \alpha^{\sigma_x} + \alpha}{2}  \right)  = 0   , \\[2ex]
\cos \frac{ \alpha^{\sigma_x} - \alpha}{2} \left(  \delta^{\Delta_y}  -  \frac{1}{2}   \sin \frac{ \alpha^{\sigma_x} + \alpha}{2}  \right)  = 0   ,
\ea \ee 
hence
\be
 \delta^{\Delta_y}  =  \frac{1}{2}    \sin \frac{ \alpha^{\sigma_x} + \alpha}{2} ,
\ee
and, substituting  $\delta = \frac{1}{2} \alpha^{\Delta_x} $,  $\alpha = \phi$, we obtain
\be
 \alpha^{\Delta_x \Delta_y}  =    \sin \frac{ \alpha^{\sigma_x} + \alpha}{2}   \qquad  \longrightarrow  \qquad   \phi^{\Delta_x \Delta_y}  =    \sin \frac{ \phi^{\sigma_x} + \phi}{2} ,
\ee
which can be considered as a special case of \rf{dsG}.

In the third case we assume $\mu_x = 0$.  Now the Lax pair essentially simplifies 
\be
\Psi^{\Delta_x} = (i \zeta \sigma_3 - i \sigma_2 \delta  ) \Psi \ , \qquad  
\Psi^{\Delta_y} =   \frac{1}{4 i \zeta} \left( \sigma_3   \cos\alpha  + \sigma_1   \sin\alpha  \right)  \Psi 
\ee
Compatibility conditions yield three equations (the term proportional to the unit matix  vanishes):
\be \ba{l}
\delta^{\Delta_y} = \frac{1}{2} \sin\alpha \ ,  \\[1ex]
  ( - \alpha^{\Delta_x} + \delta^{\sigma_y} + \delta )  \cos\alpha = 0 \ , \\[1ex]
  (  \alpha^{\Delta_x} - \delta^{\sigma_y} -  \delta )  \sin\alpha = 0 \ , 
\ea \ee
The last two equations are identically satisfied by virtue of Lemma~\ref{lem-aldel}. Then, using \rf{aldel}, we obtain
\be
     \phi^{\Delta_x \Delta_y} = \sin \frac{\phi^{\sigma_y} + \phi}{2} \ ,
\ee
which can be considered as a special case of \rf{dsG}.
\end{Proof}

\begin{cor}
The equation \rf{dsG} will be called the sine-Gordon equation on time scales. For $\mu_x \neq 0, \mu_y \neq 0$ it coincides with the Hirota equation (doubly discrete sine-Gordon equation) \cite{Hir,Or} and for $\mu_y = 0$  it coincides with the discrete sine-Gordon equation \cite{PL}. 
\end{cor}

\no The equation \rf{dsG} can be rewritten as
\be
  \phi^{\Delta_1 \Delta_2} =  F \left( \frac{1}{4} \mu_x \mu_y  \sin \langle\langle \phi \rangle\rangle \right) \sin \langle\langle \phi \rangle\rangle ,
\ee
where  
\be
   \langle\langle \phi \rangle\rangle  = \frac{1}{4} \left(  \phi^{\sigma_x \sigma_y} + \phi^{\sigma_x} + \phi^{\sigma_y} + \phi  \right) ,
\ee
and \  $F (x) := x^{-1} \arcsin x$,   \  $F (0) = 1$.

\section{Darboux-B\"acklund transformation}

The Darboux-B\"acklund transformation on time scales was formulated and derived in \cite{Ci-pseudo}.  This is a gauge-like transformation preserving the structure of the Lax pair,  see  \cite{Ci-dbt}. 
Let \  $\tilde \Psi = D \Psi$  \ ($D$ is called ``Darboux matrix''). Then  matrices  $U, V$ of the Lax pair \rf{Lax-time} transform as follows
\be  \ba{l}
\tilde U = D^{\Delta_1} D^{-1} + D^{\sigma_1} U D^{-1} \ , \\[1ex] 
 \tilde V = D^{\Delta_2} D^{-1} + D^{\sigma_2} V D^{-1} \ . 
\ea \ee
In this paper we confine ourselves to the binary Darboux matrix which adds (as a kind of nonlinear superposition) a soliton onto a given background. Fortunatelly, the Darboux matrix has exactly the same form in all cases (continuous, discrete and time scales) \cite{Ci-pseudo}:
\be  \label{Dar-mat}
\ \    D = \frac{  \lambda -  \kappa_1 A  }{\lambda - i \kappa_1}  \ .
\ee
where $A = i (I - 2 P) $  and  $P$ is a Hermitean projector ($P^2=P$, $P^\dagger = P$) defined by 
\be  \label{kerimp}
\ker P = \Psi (i \kappa_1) {\vec c}_1 \ , \quad {\rm Im}\,P = \Psi (- i \kappa_1) {\vec c}_1^\perp  \ ,
\ee
where ${\vec c}_1$ is a constant vector.  Moreover, the Lax pair \rf{Lax-time} satisfies: 
\be   \label{UV-loop}
   U (-\zeta) = \e_2^{-1} U (\zeta) \e_2 , \qquad    V (-\zeta) = \e_2^{-1} V (\zeta) \e_2 ,
\ee 
which implies the following constraint on the projector $P$ \cite{Ci-pseudo}:
\be
  \e_2 P = (I - P) \e_2 .
\ee
Hence \cite{Ci-pseudo}:
\be
P = \frac{1}{2} (I + \sigma_1 \sin\theta + \sigma_3 \cos\theta ) = \frac{1}{2} \mm 1 + \cos\theta & \sin\theta \\ \sin\theta  & 1-\cos\theta \ema  .
\ee
Taking into account well known trigonometric identities, we can represent $P$ as
\be
  P =   \m  \cos \frac{\theta}{2} \\[1ex] \sin \frac{\theta}{2} \ema \mm \cos\frac{\theta}{2} \  , & \sin \frac{\theta}{2} \ema .
\ee
Likewise, we have
\be
 I -  P =  \frac{1}{2} \mm 1 - \cos\theta & - \sin\theta \\ - \sin\theta  & 1+\cos\theta \ema   ,
\ee
and
\be
I- P =  \m - \sin \frac{\theta}{2} \\[1ex]  \cos \frac{\theta}{2} \ema \mm - \sin\frac{\theta}{2} \  , & \cos \frac{\theta}{2} \ema .
\ee
Therefore,
\be  \label{kerimp-fg}
  {\rm im} P = f   \m  \cos \frac{\theta}{2} \\[1ex] \sin \frac{\theta}{2} \ema ,
 \quad {\ker P} = g \m - \sin \frac{\theta}{2} \\[1ex]  \cos \frac{\theta}{2} \ema ,
\ee
where $f, g$ are  arbitrary functions. The function $\theta$, parameterizing the projector, can be uniquely derived from \rf{kerimp}, provided that  $\Psi (\zeta)$ is known. Alternatively, we can solve dynamic equations for $\theta$.

\ods

\begin{prop}  \label{prop-dynth}
 Dynamic  equations for $\theta$  parameterizing  the projector $P$ read
\be  \ba{l}   \label{dyn-th}
\frac{1}{\mu_x} \sin \left( \mu_x \delta - \frac{\theta^{\sigma_x} - \theta}{2} \right) = \kappa_1 \sin \frac{\theta^{\sigma_x} + \theta}{2}    ,  \qquad  ({\rm for} \ \ \mu_x \neq 0) ,   \\[2ex] 
 \frac{1}{\mu_y} \sin \frac{\theta^{\sigma_y} - \theta }{2} = \frac{1}{4 \kappa_1} \sin \left(   \alpha -  \frac{\theta +  \theta^{\sigma_y}}{2}   \right)  \qquad ({\rm for} \ \ \mu_y \neq 0)  , \\[2ex]
  \delta - \frac{1}{2} \theta,_x = \kappa_1 \sin \frac{\theta^{\sigma_x} + \theta}{2} , \qquad ({\rm for} \ \ \mu_x = 0) ,   \\[2ex]
  \theta,_y = \frac{1}{2 \kappa_1}   \sin (\alpha - \theta)  ,  \qquad ({\rm for} \ \ \mu_y = 0) .
\ea  \ee
\end{prop}

\begin{Proof} We confine ourselves to the case $\mu_x \neq 0$, $\mu_y \neq 0$. Other cases can be considered in exactly the same way (all intermediate steps can be obtain as a formal limit).   From \rf{kerimp} it follows that
\be  \label{kerdif}
    ({\ker  P})^{\Delta_x} = U (i \kappa_1) {\rm ker} P , \qquad  ({\ker  P})^{\Delta_y} = V (i \kappa_1) {\rm im} P . 
\ee
Analogical equations for ${\rm im} P$ can be omitted because in this case ${\rm im} P$ is uniquely detemined by ${\rm ker} P$. Indeed, ${\rm im P} = ({\rm ker} P)^\perp$ and ${\rm im} P = \e_2 {\rm ker} P$. Note that $ U (i \kappa_1) $ and $V (i \kappa_1)$ are real matrices, provided that all coefficients and fields are real:
\be \ba{l}
U (i\kappa_1) =   \kappa_1 \mm - 1 & 0 \\ 0 & 1 \ema  + \frac{\sin\mu_x  \delta}{\mu_x}  \mm 0 & -1 \\ 1 & 0 \ema  -  \frac{1 - \cos\mu_x \delta}{\mu_x}  \mm 1 & 0 \\ 0 & 1 \ema ,  \\[3ex]
V (i\kappa_1)  = - \frac{1}{4 \kappa_1} \mm \cos\alpha & \sin\alpha \\ \sin\alpha & - \cos\alpha \ema .  
\ea \ee
Substituting \rf{kerimp-fg} into \rf{kerdif} and dividing by, respectively, $f$ and $g$, we obtain:
\be \ba{l}  \label{four}
 (\sin  \frac{\theta}{2} )^{\Delta_x} + \frac{f^{\Delta_x}}{f} \sin  \frac{\theta^{\sigma_x}}{2}  = - \kappa_1 \sin \frac{\theta}{2}  +  \frac{\sin \mu_x \delta}{\mu_x}  \cos \frac{\theta}{2} - \frac{1 - \cos  \mu_x \delta}{\mu_x}  \sin \frac{\theta}{2}  , \\[2ex]
(\cos \frac{\theta}{2} )^{\Delta_x} + \frac{f^{\Delta_x}}{f}  \cos\frac{\theta^{\sigma_x}}{2}  = \kappa_1 \cos \frac{\theta}{2}  -  \frac{\sin \mu_x \delta}{\mu_x}  \sin \frac{\theta}{2} - \frac{1 - \cos  \mu_x \delta}{\mu_x}  \cos \frac{\theta}{2} \ ,
\\[2ex]    
(\sin  \frac{\theta}{2} )^{\Delta_y} + \frac{g^{\Delta_y}}{g} \sin  \frac{\theta^{\sigma_y}}{2}  =  - \frac{1}{4 \kappa_1}  \left(   \cos\alpha \cos \frac{\theta}{2}  -   \sin\alpha \cos \frac{\theta}{2} \right)  , \\[2ex]
(\cos \frac{\theta}{2} )^{\Delta_y} + \frac{g^{\Delta_y}}{g}  \cos\frac{\theta^{\sigma_y}}{2}  = \frac{1}{4 \kappa_1}  \left(   \sin\alpha \sin \frac{\theta}{2}  +   \cos\alpha \cos \frac{\theta}{2} \right)  , 
\ea \ee
Eliminating $f$ and $g$ from \rf{four} we get two equations
\[ \ba{l}
\frac{1}{\mu_x} \sin \frac{\theta^{\sigma_x} - \theta}{2} = - 
\kappa_1 \sin \frac{ \theta^{\sigma_x} + \theta }{2} + \frac{ \sin\mu_x \delta}{\mu_x}  \cos \frac{ \theta^{\sigma_x} - \theta }{2}  + \frac{1 - \cos\mu_x \delta}{\mu_x} \sin \frac{ \theta^{\sigma_x} - \theta }{2} , \\[2ex]
 \frac{1}{\mu_y} \sin \frac{\theta - \theta^{\sigma_y}}{2} = \frac{1}{4 \kappa_1} \sin \left(  \frac{\theta +  \theta^{\sigma_y}}{2}  - \alpha  \right)  
\ea \]
and, after elementary calculation, we get \rf{dyn-th}.    \end{Proof}

\begin{prop}[\cite{Ci-pseudo}]   \label{prop-DBT}

If $D$ is given by \rf{Dar-mat}, \rf{kerimp} and the Lax pair is of the form  
\be  \label{Lax-uv}
U = u_0 + \zeta u_1  \ , \quad V = v_0 + \frac{1}{\zeta} v_1 \ ,
\ee
where $u_0, u_1, v_0, v_1$  are linear combinations of $I, \e_1, \e_2, \e_3$ with real coefficients 
and \rf{UV-loop} is satisfied, then the Darboux-B\"acklund transformation is given by
\be \ba{l}    \label{dbt-uv}
{\tilde u}_1 = u_1 \ , \qquad 
{\tilde u}_0 = u_0 + \kappa_1 \big( u_1  A  - A^{\sigma_x} u_1 \big) \ , \\[2ex]  \displaystyle
{\tilde v}_0 = v_0 \ , \qquad   \dis 
{\tilde v}_1 = - A^{\sigma_y}  v_1  A^{-1}  \ ,
\ea \ee
where \ $A = i (I-2P) = \e_3 \cos\theta + \e_1 \sin\theta$.  
\end{prop}

Proposition~\ref{prop-DBT} can be applied directly to the Lax pair \rf{Lax-time}.  Two equations are trivially satisfied ($\tilde u_1 = u_1 = - \e_3$ and $\tilde v_0 = v_0 = 0$).  The equation for $\tilde u_0$ splits into two equations:
\be \ba{l} \dis  \label{sys-1}
 \frac{\sin \mu_x \tilde \delta}{\mu_x} = \frac{\sin \mu_x \delta}{\mu_x} - \kappa_1 (\sin\theta^{\sigma_x} + \sin\theta) , 
\\[3ex]  \dis
\frac{1 - \cos\mu_x \tilde \delta}{\mu_x} =  \frac{1 - \cos\mu_x \delta}{\mu_x} + \kappa_1 (\cos \theta^{\sigma_x} - \cos\theta) ,
\ea \ee
for $\mu_x \neq 0$. In the case $\mu_x =0$ we have, instead, 
\be \ba{l}  \label{sys-11}
 \tilde \delta = \delta - 2 \kappa_1 \sin\theta \ , \\[1ex]
\frac{1}{2} {\tilde \delta}^2  =   \frac{1}{2}  \delta^2  - \kappa_1 \theta,_x \sin \theta . 
\ea \ee
The last equation of \rf{dbt-uv}  will be simplified using an elegant approach of Clifford numbers instead of cumbersome straightforward calculations. 

\begin{lem}   \label{lem-prop}
For any scalar function $\beta = \beta(x,y)$ we have
\be \ba{l}
  \cos\beta  + \e_1 \e_3 \sin\beta = \exp (\beta \e_1 \e_3)  , \\[1ex]
  \e_3 \cos\beta + \e_1 \sin\beta = \e_3 (\cos\beta  + \e_1 \e_3 \sin\beta) = \e_3 \exp (\beta \e_1 \e_3) , \\[1ex]
  \e_3 \exp (\beta \e_1 \e_3) =   \exp (- \beta \e_1 \e_3)  \e_3 
\ea \ee
\end{lem}

\begin{Proof}   We use \rf{Cliff} and observe that $(\e_1 \e_3)^2 = - 1$.  The first equation can be proved using power series, like the famous formula  $\cos \beta + i \sin\beta = \exp (i \beta)$. 
\end{Proof}

Taking into account properties listed in Lemma~\ref{lem-prop} we can easily transform the last equation of \rf{dbt-uv} into 
\be
e_3 \exp (\tilde \alpha \e_1 \e_3) = - e_3^3 \exp (\e_1\e_3 \theta^{\sigma_x}) \exp (- \e_1\e_3 \alpha) \exp (\e_1\e_3 \theta) .
\ee
Because $\e_3^3 = - \e_3$ and exponents of all exponential functions commute, we have
\be  \label{eq-al}
  \e_3 \exp(\tilde \alpha \e_1 \e_3) = \exp \e_1 \e_3 (\theta^{\sigma_x} + \theta - \alpha) ,
\ee
hence  we finally get:  \ $\tilde\alpha = - \alpha + \theta^{\sigma_x} + \theta$.  
\ods

\begin{prop}
The Darboux-B\"acklund transformation \rf{dbt-uv} for the Lax pair \rf{Lax-time} is equivalent to the system:
\be \ba{l}   \label{sys-aldel}
\tilde\alpha = - \alpha + \theta^{\sigma_y} + \theta ,  \\[1ex]
\tilde \delta  + \delta = \theta^{\Delta_x} , 
\ea \ee
which can be reduced to one  equation for $\phi$:
\be \ba{l}  \label{sys-phi}
    \tilde \phi = 2 \theta - \phi \ ,
\ea \ee
where $\theta$ is computed either from \rf{kerimp} or from Proposition~\ref{prop-dynth}. 
\end{prop}

\begin{Proof}  The first equation of \rf{sys-aldel} was already obtained from \rf{eq-al}. Considering the second equation,  we first  assume $\mu_x \neq 0$. The system \rf{sys-1}  can be rewritten as
\be \ba{l}
\frac{1}{\mu_x} \sin \frac{\mu_x (\tilde \delta - \delta)}{2} \cos  \frac{\mu_x (\tilde \delta + \delta)}{2} = -  \kappa_1 
\sin \frac{\theta^{\sigma_x} + \theta}{2} \cos  \frac{\theta^{\sigma_x} - \theta}{2} , \\[2ex]
\frac{1}{\mu_x} \sin \frac{\mu_x (\tilde \delta - \delta)}{2} \sin  \frac{\mu_x (\tilde \delta + \delta)}{2} = -  \kappa_1 
\sin \frac{\theta^{\sigma_x} + \theta}{2} \sin  \frac{\theta^{\sigma_x} - \theta}{2} , 
\ea \ee
Dividing these equations side by side, we get
\be \ba{l}
  \frac{1}{\mu_x} \tan   \frac{\mu_x (\tilde \delta + \delta)}{2} = \  \frac{1}{\mu_x}  \tan \frac{\theta^{\sigma_x} - \theta}{2} ,
\ea \ee
which implies the first equation of  \rf{sys-aldel}. The remaining equation,
\be \ba{l}
 \frac{1}{\mu_x}  \sin \frac{\mu_x (\tilde \delta - \delta)}{2} = - \kappa_1 \sin \frac{\theta^{\sigma_x} + \theta}{2} ,
\ea \ee
after substituting $\tilde \delta = \theta^{\Delta_x} - \delta$,  reduces to the first equation of \rf{dyn-th}.  
The case $\mu_x = 0$  is  simpler.  Substituting $\tilde \delta - \delta$ from the first equation of  \rf{sys-11} into the second equation of \rf{sys-11} we get the second  equation of \rf{sys-aldel}.

Finally, substituting \rf{aldel} into \rf{sys-aldel}, we get 
\be  \ba{l}
 \left(  \tilde \phi + \phi - 2 \theta \right)^{\Delta_x} = 0 \ ,   \\[2ex]
 \left(  \tilde \phi + \phi - 2 \theta \right)^{\sigma_y} =  -  ( \tilde \phi + \phi - 2 \theta)   \ . 
\ea \ee
Hence, $\tilde \phi + \phi - 2 \theta = \chi$, where $\chi$ is a function of $y$  such that  $\chi^{\sigma_y} = - \chi$. Without loss of generality we can take $\chi=0$ because transformation $\phi \rightarrow \phi + \chi$  does not change the Lax pair \rf{Lax-time}. 
\end{Proof}

\section{Solitons  on an arbitrary time scale}

Pure soliton solutions are obtained, as usual, by applying the Darboux-B\"acklund  transformation to the simplest (``vacuum'')  background $\phi = 0$.  In this case the Lax pair reduces to 
\be
 \Psi^{\Delta_x} = i \zeta \sigma_3 \Psi \ , \qquad  \Psi^{\Delta_y} = \frac{1}{4i \zeta} \sigma_3 \Psi ,
\ee
which can be easily solved (assuming initial condition $\Psi (0,0) = I$): 
\be  \label{triv-lax}
  \Psi = \mm e_{i\zeta} (x) e_{\frac{1}{4i \zeta}} (y) & 0 \\[2ex] 0 &  e_{- i\zeta} (x) e_{- \frac{1}{4i \zeta}} (y) \ema ,
\ee
where $e$ denotes the delta exponential function \rf{deltaexp}.   
From the previous section we know that $\tilde \phi = 2 \theta - \phi$. Therefore 
one-soliton solution is given by
\be
       \phi_{sol} = 2  \theta ,
\ee
where $\theta$ is associated with  the simplified linear system \rf{triv-lax}.  

\begin{prop}
One-soliton solution to the sine-Gordon equation on an arbitrary time scale (equation \rf{dsG})  is given by
\be   \label{onesol}
\phi_{sol} = 4 \arctan \left( c \ E_{- 2 \kappa_1} (x) E_{ - \frac{1}{2 \kappa_1}} (y) \right) ,
\ee
where $E$ denotes the  Cayley-exponential function \rf{Cayley}  and  $c, \kappa_1$ are real parameters. 
\end{prop}

\begin{Proof}  
In the case $\alpha=\delta= 0$ equations \rf{dyn-th} read
\be  \ba{l}  \label{4-eqs}
\frac{1}{\mu_x} \sin \frac{\theta^{\sigma_x} - \theta}{2} = - \kappa_1 \sin \frac{\theta^{\sigma_x} + \theta}{2} ,
 \qquad  ({\rm for} \ \  \mu_x \neq 0)   \\[2ex]
\frac{1}{\mu_y} \sin \frac{\theta^{\sigma_y} - \theta}{2} = - \frac{1}{4 \kappa_1}  \sin \frac{\theta^{\sigma_y} + \theta}{2} ,    \qquad  ({\rm for} \ \  \mu_y \neq 0)
\\[2ex]
\theta,_x = - 2 \kappa_1 \sin  \theta ,  \qquad  ({\rm for} \ \  \mu_x =0)  ,   \\[2ex]
\theta,_y = - \frac{1}{2 \kappa_1} \sin\theta \qquad  ({\rm for} \ \  \mu_y =0) .
\ea \ee
Equations \rf{4-eqs} can be easily transformed into:
\be \ba{l}
   \tan \frac{\theta^{\sigma_x}}{2} =   \frac{1 - \mu_x \kappa_1}{1+\mu_x \kappa_1}  \tan \frac{\theta}{2} , 
 \qquad  ({\rm for} \ \  \mu_x \neq 0) \\[3ex]
     \tan \frac{\theta^{\sigma_y}}{2} =   \frac{1 - \frac{\mu_y}{4 \kappa_1}}{1+\frac{\mu_y}{4 \kappa_1}}  \tan \frac{\theta}{2} , 
 \qquad  ({\rm for} \ \  \mu_y \neq 0)   \\[3ex]
 \left(\tan \frac{\theta}{2} \right),_x = - 2 \kappa_1   \tan \frac{\theta}{2} \ ,    \qquad  ({\rm for} \ \  \mu_x =0)  ,   \\[3ex]
\left(\tan \frac{\theta}{2} \right),_y = - \frac{1}{2 \kappa_1}   \tan \frac{\theta}{2} \ ,    \qquad  ({\rm for} \ \  \mu_y =0)
\ea \ee
which yields \rf{onesol} by virtue of \rf{Cayley} (the continuous case agrees with this solution, as well).
Another method to get \rf{onesol}, even more straightforward, consists in using  \rf{kerimp} and \rf{kerimp-fg}.
\end{Proof}

In the continuous and discrete case the formula \rf{onesol} reduces to known expressions:
\be  \ba{l}
   \phi_{sol-cont} = 4 \arctan \exp (- 2 \kappa_1 x  - \frac{1}{2 \kappa_1} y + c_0 ) \ , \\[2ex]
 \phi_{sol-discr} =  4 \arctan \left(  c  \left( \frac{1 - \kappa_1 h_x}{1 + \kappa_1 h_x} \right)^{\frac{x}{h_x}} 
   \left( \frac{1 - \frac{h_y}{4 \kappa_1} }{1 + \frac{h_y}{4 \kappa_1}} \right)^{\frac{y}{h_y}}    \right) ,
\ea \ee
where $h_x=\mu_x$ and $h_y=\mu_y$. The discrete soliton has, in fact, a similar shape as the smooth soliton. Indeed,
\be
  \phi_{sol-discr} =  4 \arctan \exp (- 2 \tilde\kappa_1 x  - \frac{1}{2 \hat\kappa_1} y + c_0 ) ,
\ee
where $c = \exp c_0$ \  and
\be
\tilde \kappa_1  = \frac{1}{2 h_x} \ln \left( \frac{1 + \kappa_1 h_x}{1 - \kappa_1 h_x} \right) , \qquad 
\hat \kappa_1^{-1}  = \frac{1}{2 h_y} \ln \left( \frac{1 + \frac{h_y}{4 \kappa_1}}{1 - \frac{h_y}{4 \kappa_1}} \right) .
\ee
Therefore, on two most popular time scales ($\T=\R$, $\T= h \Z$) solitons are solitary waves. In the discrete case  velocity and shape depend on $h_x$ and $h_y$. However, in general, one-soliton solution \rf{onesol} is not a solitary wave. 

\section{Summary}

We formulated and studied an integrable analogue of the sine-Gordon equation on an arbitrary time scale. In particular, we provided parallel computations for the discrete and continuous case. The Darboux-B\"acklund transformation has the same form on any time scale. Soliton solutions of the sine-Gordon equation on time scales are expressed by the Cayely exponential function. Therefore, like in the continuous case, soliton solutions of the nonlinear equation are expressed in terms of exponential functions which were introduced as solutions to linear equations.   

\ods

{\it Acknowledgement.} The first author (J.L.C.) is partly supported  by the National Science Centre (NCN) grant no. 2011/01/B/ST1/05137.

\end{document}